\documentstyle[preprint,prl,aps]{revtex}
\gdef\journal#1, #2, #3, 1#4#5#6{           
{\sl #1~}{\bf #2}, #3 (1#4#5#6)}            

\def\anyas{\journal Ann. N.Y. Acad. Sci., }

\def\epl{\journal Europhys. Lett., }

\def\jcp{\journal J. Chem. Phys., }

\def\jpc{\journal J. Phys. Chem., }

\def\jpp{\journal J. Phys. (Paris), }

\def\jppii{\journal J. Phys. II (Paris), }

\def\mac{\journal Ma\-cro\-mo\-le\-cu\-les, }

\def\nat{\journal Nature, }

\def\pnas{\journal Proc. Natl. Acad. Sci. USA, }

\def\pra{\journal Phys. Rev. A, }

\def\prl{\journal Phys. Rev. Lett., }

\def\sci{\journal Science, }

\input {epsf.tex}
\begin{document}
\draft
\title{Membranes in rod solutions:\\ a system with spontaneously
broken symmetry}
\author{K. Yaman and P. Pincus}
\address{Department of Physics, University of California, Santa Barbara,
California 93106--9530}
\author{C. M. Marques}
\address{R.P.-C.N.R.S., Complex Fluid Laboratory UMR166, Cranbury,
 NJ 08512--7500, USA}
\date{\today}
\maketitle
\begin{abstract}
We consider a  dilute  solution of infinitely rigid rods near
a curved, perfectly repulsive  surface  and study the contribution of the
rod depletion layer  to the bending elastic constants of membranes. We
find that a
spontaneous curvature state can be induced
by exposure of {\sl both} sides of the membrane to a rod solution. A similar
result applies for rigid disks with a diameter equal to the rod's
length. We also
study the confinement of rods in spherical and cylindrical
repulsive shells. This helps elucidate a recent discussion on
curvature effects in confined quantum mechanical and  polymer systems.
\end{abstract}
\narrowtext

The  elastic properties of fluid membranes are among the ruling factors
in the physics of many biological and surfactant systems such as cell
membranes, vesicles, lyotropic liquid crystals or  microemulsions~\cite{safr}.
When  macromolecular species are also present in such systems, the
interactions between the membranes and the macromolecules change the bare
elastic constants. Recent studies have explored
contributions from adsorption~\cite{dege,broo1,broo2,podg,eisen},
depletion~\cite{broo2,podg,eisen}
or end-grafting~\cite{podg,cant,miln,lipo} of
flexible polymers. In all of these cases,
exposure of both sides of the membrane to a
solution results in a modification of the bending and splay modulii $K$ and
$\bar K$, the sign of the contribution depending on the possibility of
surface-bulk equilibration. For instance, depleted or adsorbed polymers reduce
$K$ and increase  $\bar K$, whereas the reverse holds for end-grafted
polymers.

Interest in solutions  of rigid macromolecules can be traced to the
extraction of tobacco mosaic virus (TMV)~\cite{stan} and to the subsequent
observation of the nematic phase in TMV solutions~\cite{bawd}, later explained
by the seminal  theory of Onsager~\cite{onsa}.  Rod shaped particles in the
colloidal range have since been studied in a large variety of mineral and
organic systems~\cite{buin}, they  are also present in the biological realm
where examples  range from  TMV-like virus  to fibrils of amyloid
$\beta$-protein, the molecular agent at the origin of the Alzheimer disease.
The surface interactions  of rigid macromolecules were
first studied by Asakura and Oosawa~\cite{asak} who showed that the steric
depletion of the molecules at a flat surface increases the interfacial
energy of
the system, implying that  two surfaces brought to  separations smaller than a
rod length will experience an attractive force.  The theory of depletion
interactions for rods  have since been extended  to include
ordering effects of the bulk phase~\cite{poni} or effects of  rod-rod excluded
volume interactions~\cite{mao}; but it has been mostly~\cite{auvr}
devoted to flat
geometries. In this letter we  extend  the original Asakura and Oosawa
theory  to
include effects of surface curvature and determine the contribution of the rod
depletion layer to the elastic constants of the membranes.  An important reason
for considering this case  can be   better understood  by comparing
the order of
magnitudes of changes in $K$ and $\bar K$ caused by depleted layers of
rods with those caused by spherical
particles. In a solution of  colloidal spheres of radius $r_0$ and particle
number density $\rho_b$, the typical scale of the  energy density is
$k_B T \rho_b$. Corrections to the interfacial energy are thus of the order
$\Delta \gamma \simeq k_B T \rho_b r_0$. On other hand, interfacial tensions in
most liquids are of the order of $\gamma_0\simeq k_B T/a^2$ where $a$ is a
microscopic size. For instance for $a\sim 0.1$nm, $\gamma_0$  is of  the
order of tens of mN/m. The corrections due to the depletion of spherical
particles  are thus a factor
$(a/r_0)^2$ lower than typical interfacial tensions even at order unity volume
fractions $\phi = \rho_b (4 \pi/3) r_0^3$.  Curvature corrections to the
interfacial energy can generally be written as $\Delta \gamma \simeq k_B T
\rho_b
r_0 ( 1 + C_1 r_0/R + C_2 r_0^2/R^2) $, with $R$ the radius of curvature and
$C_1, C_2$ two  numerical constants. This leads to modifications
of the bare  elastic constants
of the order of $\Delta K \simeq k_B T \rho_b r_0^3$: at
the upper concentration limit $\rho_b\sim 1/r_0^3$, these corrections are
only of
order $k_B T$, a value at the lower end of the range\cite{something},
$1-20 k_B T$, of most bare
elastic constants. In the case of a rod solution with rod number
density $\rho_b$,  the upper concentration limit of the isotropic solution is
the Onsager concentration $\rho_b^\star=4.2/(L^2 D)$ \cite{onsa},
with $L$ the length of the
rod and $D$ its diameter.
The contributions to the interfacial tension are now of
the order   $\Delta \gamma \simeq k_B T \rho_b L$, but even for  rod
diameters of
order of the microscopic length $a$  the contribution to the
interfacial tension
of rod solutions at the Onsager concentration is still a factor ($a/L$) smaller
than typical interfacial tension values. However, modifications of the elastic
constants are here of order $\Delta K \simeq k_B T \rho_b L^3$, a factor
$(L/D)$ larger than $k_B T$.
Therefore, even rather rigid phospholipid membranes with  elastic
constants as large as 20$k_B T$ may have their rigidities modified
substantially at low rod concentrations where rod-rod interactions are
negligible. We remark also that the steric surface interactions
 are identical for rigid rods and rigid disks  if one takes
the rod's length and the disk's diameter to be the same; however the disk
system does not enjoy the above-mentioned $L/D$ enhancement.

We consider an ideal gas of rods of length $L$ in the presence of
flat and curved surfaces which repel the rods. We parameterize the possible rod
configurations by the center of mass coordinates,
${\vec{r}}$, and the two angles specifying in which direction  the rod points,
$\omega \equiv (\theta, \phi)$. For sake of simplicity the
thickness of the rods is taken to be zero. The relevant
potential describing the
thermodynamics of the system can be written as
\begin{eqnarray}\label{om}
\Omega\left[\rho({\vec{r}}, \omega)\right] =
\int d\,{\vec{r}} \int d\, \omega \,\rho({\vec{r}}, \omega)
\left[\log(v \, \rho({\vec{r}}, \omega)/e) - \right. \nonumber \\
\left. (\mu_b - U_{ext} ({\vec{r}}, \omega)) \right]
\end{eqnarray} 
where $v$ is some normalization volume,
$\mu_b$ is the solution chemical potential, and  $U_{ext}$ is the
hard wall
interaction potential that is either infinite or zero, depending on
whether the configuration of the rod is allowed by the `hard wall' requirement
or not. Functional minimization of $\Omega$
with respect to $\rho({\vec{r}},\omega)$
gives the equilibrium density profile:
\begin{equation}  
\rho({\vec{r}}, \omega)  =
{e^\mu \over v} e^{-U_{ext} ({\vec{r}}, \omega)}
        \equiv \frac{\rho_b}{4 \pi} e^{-U_{ext} ({\vec{r}}, \omega)}
\label{rhobar}
\end{equation}  
The local, position dependent, number density of particles $\rho({\vec{r}})$ is
simply the sum over all allowed angular configurations of $\rho({\vec{r}},
\omega)$. The excess surface energy is then calculated from
\begin{eqnarray}
\label{delta} 
\Delta \gamma ={\Omega\left[\rho({\vec{r}}, \omega)\right]
-\Omega\left[\rho(z\to \infty, \omega)\right]\over S} = \nonumber\\
\int d\,z \, (\rho_b-\rho(z)) \,J(z,R)
\end{eqnarray} 
where $S$ is the surface area, $z$ the perpendicular distance from the
surface and $J(z,R)$ a phase space factor, which depends on the
geometry. For flat surfaces $J(z,R)=1$, for cylinders of radius R,
$J(z,R)=1+z/R$, and for spheres $J(z,R)=(1+z/R)^2$.  From
equation~(\ref{delta})
it is clear that differences between the excess energy of a flat and a curved
surface depend on two factors. The first is the configurational part
measured by
the differences in the rod density profiles, the second one is associated with
the space available to  the center of mass in the neighborhood of the
surface, and
is measured by $J(z,R)$. In the case of hard sphere solutions, where
there is no
coupling between configuration and curvature, only $J(z,R)$ is responsible for
energy differences; ($\rho(z) = \rho_b \, \Theta(z - L/2)$, independent
of geometry).
It is easy to show that in that case one has
$\Delta \gamma = k_B T \rho_b r_0 ( 1 + r_0/(2 R))$
for cylindrical surfaces and
$\Delta \gamma = k_B T \rho_b r_0 ( 1 + r_0/R + r_0^2/(3 R^2))$
for spherical ones. Following
the usual procedure~\cite{safr} one then finds the corrections to the elastic
constants of a membrane exposed to a colloidal solution: $\Delta K =0$ and
$\Delta \bar K = 2/3 \,k_B T \, \rho_b r_0^3$.

Figure 1 shows the angular geometrical constraints for a rod close to a
flat or curved surface. When the center of mass is at a perpendicular
distance larger than
$L/2$ from the surfaces, the direction of the rod is not
constrained. For
distances $0<z<L/2$ only a fraction of the  solid angle is available
to the rod.
For rods outside a curved geometry a second distance  $z_c^o$ must be defined:
for $z_c^o<z<L/2$ the rod touches the surface with its end, whereas for
distances
smaller than $z_c^o$ the rod touches the surface somewhere along its length.
Inside a curved geometry yet a different length $z_c^i$ must be defined, below
which the rod is geometrically not allowed. Performing the angular
integrations
one gets the density profiles plotted in figure 2. As expected rods outside a
spherical curved  surface have a larger density than rods  close to a flat
surface. Conversely, rods inside a spherical surface have a smaller density
than
rods  close to the flat surface.  The density profile of rods close to
cylindrically bent surfaces is identical to the spherical profile when the rods
are oriented perpendicularly to the cylinder axis, and to the flat
profile when the
rods are oriented along the cylinder's axis. Intermediate angles interpolate
smoothly between the two limits. A notable feature of the density profiles is
that they are non-analytic functions of the curvature. For instance,
the inner
part ($z<z_c^o$) of the external profile depends on the square root of the
curvature, and the internal profile has a cutoff at $z_c^i$. This
implies  that even for large radii of curvature one cannot simply
obtain the internal profiles from the external one by a $R\to -R$
transformation.
Integration of the profiles lead to the following contributions to the
interfacial energies:
\begin{eqnarray}\label{delout}
\Delta \gamma_{\rm out} =k_b T \rho_b
{L\over 4}
\end{eqnarray}
for rods outside spheres
and  cylinders, or close to flat surfaces;
\begin{eqnarray}\label{delin}
\Delta\gamma_{\rm in} =k_b T \rho_b {L\over 4} \,
\left(1 - \alpha {L^2 \over R^2}\right)
\end{eqnarray} 
for rods inside spheres
($\alpha = 1/12$) and  cylinders ($\alpha = 1/32$). Results for the outside
configuration are exact.
For the inside configuration
results are exact for spheres, and perturbative to second
order in $1/R$, for cylinders.
Equations~(\ref{delout}) and (\ref{delin}) bear a few
interesting consequences. For instance
equation~(\ref{delout}) implies that the excess free energy  of a
convex volume
immersed in a rod solution does not depend on how the surface of the object is
curved, but only on its area. Also, equation~(\ref{delin}) indicates that
flexible membranes which expose one surface  to a rod solution will
spontaneously
bend towards the solution. Note however that there is no spontaneous
curvature in
the usual sense of a contribution to the energy linear in $1/R$.  When the
membrane is immersed in the solution (exposure of both sides of the
membrane) the
total surface energy excess is given by the sum of the two
contributions
$\Delta \gamma_{\rm in}$ and  $\Delta \gamma_{\rm out}$.
Curving the surface still
decreases the total free energy, but in this symmetric situation there is no
preferred side toward which the membrane should bend. Immersion will therefore
spontaneously break the symmetry of the system. To the extent that the
non-analytic character of the total free-energy can be neglected
(there is a
finite discontinuity in the second derivative at $1/R=0$), the following
contributions to the membrane elastic constants can be extracted
\begin{eqnarray}\label{dkappa} 
\Delta K=- k_B T \rho_b {L^3 \over 64} = - k_B T {\rho_b\over \rho_b^\star}
{1\over 15.2}{L \over D},
\end{eqnarray}  
\begin{eqnarray}\label{dkappabar}  
\Delta \bar K
= k_B T \rho_b {L^3 \over 96} = k_B T {\rho_b\over \rho_b^\star}
{1\over 22.9}{L\over D}.
\end{eqnarray} 
One finds thus a reduction in $K$ and an increase in $\bar K$ \-- this
favors the formation of saddle structures (for instance, cubic phases
with periodic
minimal surfaces). This is similar to other depletion and equilibrium
adsorption
problems, but as explained above the amplitude of the contributions  in this
system is very  large (of order of $L/D$ times larger than $k_b T$, a
value to
be compared  for instance to $k_b T \ln N$ for adsorbed polymers with
polymerization index $N$).

In a confined geometry, e.g. in ordered stacks of membranes
(lyotropic smectics), in multilamellar vesicles
or between any other two surfaces separated by some distance $d$ the rod
configurations are restricted by two constraints. If
the distance $d$ is larger than the rod length the confinement effect can be
calculated from the above results simply by summing the effects of two
non-overlapping depletion layers. If the surfaces are at a distance
smaller than the rod's length the two depletion layers overlap, and one
needs to compute two-surface effects specifically. The overlap of
the  rod depletion layers is known to lead to an attractive force between
the two
surfaces, but recent calculations that include rod-rod excluded volume
interactions in flat geometries show that a repulsive force  can set in at
distances of the order of the rod length \cite{mao}.
Our method can in principle be
applied to the evaluation of the force between two convex spherical or
cylindrical surfaces, but we postpone presentation of those results to a
forthcoming paper and concentrate here on the simpler case of confinement
in spherical or cylindrical hard shells. This problem is related to a recent
discussion of the curvature effects in confined quantum mechanical or
macromolecular systems. Indeed, if a quantum mechanical particle is confined in
a one or two dimensional manifold embedded in three dimensions, its movement
along that manifold will be described by the usual Schr\"odinger equation with
an additional curvature dependent potential. For instance a quantum particle
moving in a confinement tube will always be attracted by curved regions of that
tube. A quantum particle in a two dimensional manifold will experience an extra
potential\cite{jensen},
$U = - {\hbar^2\over 2 m} {1\over 4}
\left({1\over R_1}-{1\over R_2} \right)^2$, with $R_1$ and $R_2$
the two principal radii of curvature.
Particles will be attracted to curved regions of the surface except when it is
spherically bent ($R_1=R_2$). It has also been shown that such curvature
effects propagate into the classical world of statistical mechanics and that
the quantum mechanical results hold also for flexible polymers confined into
spherically or cylindrically bent gaps\cite{kerem}.
For such a system the difference in the free energy
of confined states in curved shells as compared to flat ones is given by
$U = - k_B T \rho_b R_G^2 {1\over 4} \left({1\over R_1}-{1\over
R_2} \right)^2$, where $R_G^2 \equiv N \, a^2$, $a$ being the monomer
size.  Apart from the typical energy scale, these potentials
are identical. This identity is due to similarities in the equations
governing the behavior of both systems, and in fact holds more generally for
other systems containing Laplacians in their equations of motion. In this
context it is also interesting to investigate the confinement effect  for
other classical systems, whose configurations  are not determined by Laplacian
equations. For confined colloidal particles it is convenient to study these
effects by considering open curved gaps, i.e. shells that allow the
concentration
of their solutions to equilibrate with an external reservoir. Attractive
curvature effects under these conditions will show up as an increase of the
average concentration in the curved gap as compared to non curved gaps of
identical thickness. For illustration  we compute first to second order in
$1/R$ the average concentration for a confined solution of dilute hard spheres:
\begin{equation}
\begin{array}{lccl}
\bar \rho_{\rm flat}= \rho_b (1- 2 {r_0\over d})\\
\bar \rho_{\rm sph}= \rho_b (1- 2 {r_0\over d}) \,
 (1 - {(d-r_0)\over 3} {r_0\over R^2})\\
\bar \rho_{\rm cyl}=  \rho_b (1- 2 {r_0\over d})
\end{array}
\end{equation}
A spherical curvature is therefore repulsive to the particles, whereas the
cylindrical bent shells behave as  flat gaps. For the confined rod-solution we
report the value of the average concentration for the flat, spherical and
cylindrical geometries when the surface separation is equal to the rod length
($d=L$)
\begin{equation}
\begin{array}{lccl}
\bar \rho_{\rm flat}(d=L)= { \rho_b\over 2}  \\
\bar \rho_{\rm sph}(d=L)=  { \rho_b\over 2} (1 - {1\over 16}{ 1\over R^2}) \\
\bar \rho_{\rm cyl}(d=L)=  { \rho_b\over 2} (1 +  {1\over 128}{ 1\over R^2})
\end{array}
\end{equation}
At this separation the spherical shell is  repulsive for the rod particles,
and the cylindrical shell is attractive. For lower distances both shells
become repulsive.  Results for several different systems are presented in
Table I.
Despite quantitative differences, it is apparent that cylindrical shells are
always more attractive than spherical shells. Comparison between the hard
spheres case  and the rod and the polymer case also indicate that
the coupling between curvature and configuration reinforces the primary effect
of volume exclusion. Note however that all the systems quoted in
the Table have a repulsive interaction with the confining
walls. Further insight
into this problem and a test for the generality of the conclusions drawn here
will require the study of systems where the particles are attracted to or bound
to the confining walls. Work under progress includes for instance end-grafted
polymers and adsorbed rod-like particles.

This work was partially supported by CNRS and NATO fellowships, and
NSF grants DMR-9624091 and MRL-DMR-9632716.
Acknowledgment is also made to the donors of the Petroleum Research Fund,
administered by the ACS, for support of this research ($\sharp$29306-AC7).

\begin{figure} [t]
\epsfxsize=6.0in
\centerline{\epsfbox{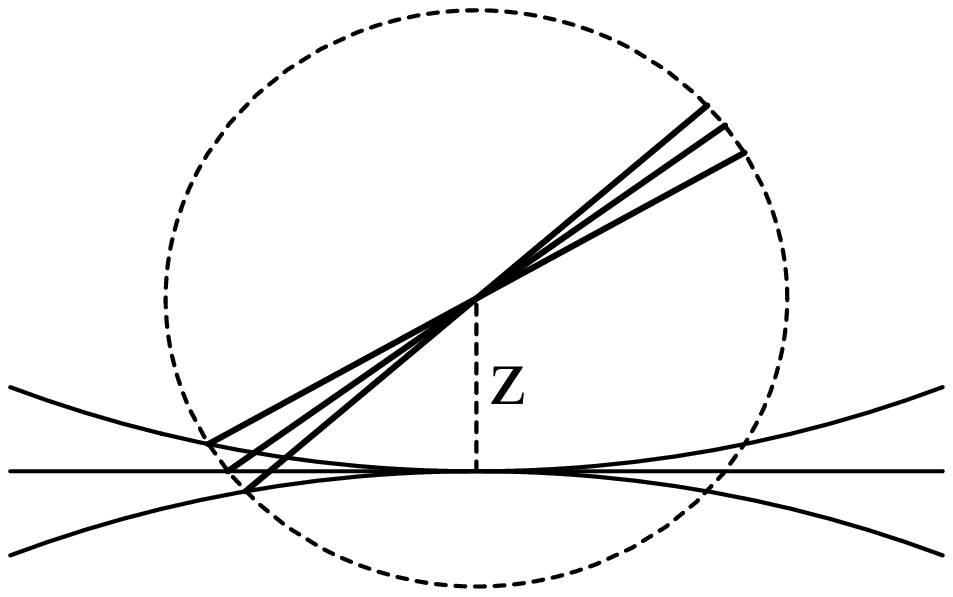}}
\caption{Configurational space of  a rod near a flat surface, outside a curved
surface, and inside of a curved surface. The rods are represented at the angle
where they touch the surface. The complete allowed angular space is
obtained by
rotation of the figure around the $z$-axis.  }
\label{rodsurf}
\end{figure}

\pagebreak

\begin{figure} [t]
\epsfxsize=6.0in
\centerline{\epsfbox{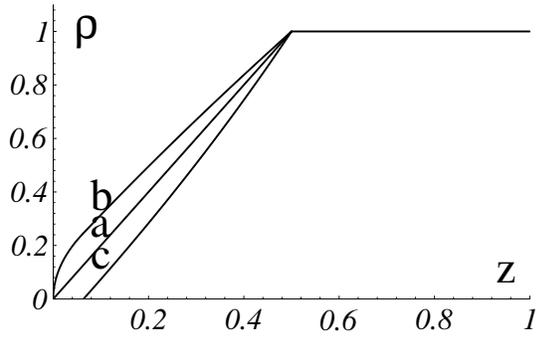}}
\caption{Profiles for the the rod number density $\rho (z)$,
in $\rho_b$ units, close to:
{\bf a)} a flat surface;
{\bf b)} outside a spherical surface; {\bf c)} inside a spherical surface.
In the particular case shown here we have $L=1$,
$R=2$, $z_c^o=0.061$ and $z_c^i=0.064$.  }
\label{profile}
\end{figure}

\pagebreak

\begin{figure}

\vskip 0.3truecm

\hrule
\vskip  0.1truecm
{System \hfill Spherical \hfill Cylindrical }

\vskip  0.1truecm
\hrule
\vskip  0.1truecm
{Quantum Particle \hfill Neutral \hfill Attraction }

{Gaussian Chain$\,\,\,\,\,\,$ \hfill Neutral \hfill Attraction }

{Hard Spheres \hfill Repulsion \hfill Neutral }

{Rods$\qquad \qquad $ \hfill Repulsion \hfill Attraction }
\vskip  0.1truecm
\hrule
\vskip 0.3truecm

\caption{Table of the  potentials induced by curvature in several  systems }
\label{table}
\end{figure}

\end{document}